\documentclass[a4paper,10pt]{article}
\usepackage{graphicx} 
\usepackage{microtype}
\usepackage{xurl}
\usepackage{authblk}
\usepackage{geometry}
\usepackage{hyperref} 
\usepackage[
    backend=biber,
    style=apa,
    natbib=true
]{biblatex}
\title{The Metastable Mind: Neural Underpinnings of Naturalistic Cognition Through the Synthesis of Event Segmentation and Metastable Neural States}

\author{Dora Gozukara}
\author{Nasir Ahmad}
\author{Djamari Oetringer}
\author{Linda Geerligs}
\affil{Donders Institute for Brain, Cognition, and Behaviour, Donders Center for Cognition, Nijmegen, Netherlands\\
e-mail: dora.gozukara@donders.ru.nl}

\begin{document}
\maketitle

\section*{Abstract}
A multitude of findings and theories from cognitive, behavioural and computational neuroscience show that neural activity unfolds in a variety of meaningful temporal units.
Behavioural research on event segmentation (ES) has shown that continuous experience is segmented into discrete events and sub-events, which aid real-time comprehension, memory, and decision-making.
Computational neuroscience research observes and models ongoing brain activity as a series of stable population activity that occur across wide spatial and temporal scales, referred to as metastable neural activity (MNA).
Through this review, we show that these isolated branches of literature, the cognitive theory of Event Segmentation (ES) and the mechanistic approach of metastability (MNA), actually study the same metastable neural states from different perspectives.
While the behavioural branch offers a theory for the cognitive and behavioural utility of segmentation, the metastability literature provides the mechanistic account at the implementational level.
We describe how metastable neural states act as the fundamental computational units of cognition and identify a number of core principles of how they operate.
One is the spatio-temporally nested hierarchy of states, where longer-duration states in higher-order regions both constrain and are shaped by  states in faster-operating regions. 
Another is that neural states are a reflection of underlying predictive models which shape perception, decision making, memory encoding and recall. 
And finally that neural states are periods of more modular processing, which are interspersed by boundaries where there is a reconfiguration of connectivity. 
Understanding how neural states emerge, interact, and shape cognition brings us closer to understanding the brain in its natural mode of operation.

\section{Introduction}
The tension between the continuous flow of our surroundings and the discrete nature of our thoughts has long been a subject of philosophical and scientific inquiry.
This tension is perhaps most famously captured in Zeno of Elea’s ``Arrow Paradox."
Zeno argued that at any single, indivisible ``now," an arrow in flight occupies a space exactly equal to its own dimensions and is, therefore, at rest.
If time is composed of such motionless instants, he reasoned, then the continuity of motion is logically impossible.
Zeno's paradox highlights our recognition that the external world seems to be ever-flowing, yet our minds impose a state-like analysis of, and architecture to, experience. 
This is most apparent when we look back at our past experiences, since we often do not experience our memories as a continuous progression of instants, but we recall distinct events that occurred.
In the context of neuroscience, we face a modern reflection of this recognition.
Rather than attempting to process every infinitesimal moment of a continuum, the brain segments experience into discrete, stable windows of activity.
However, the scientific understanding of these stable windows is currently fragmented.
Segmenting ongoing information into distinct events, is a core aspect of everyday perception and action \citep{zacks_event_2007, kurby_segmentation_2008}.
There is a broad literature on event segmentation (ES), which has provided insights into a wide range of interconnected cognitive phenomena (such as perception, attention, memory, and prediction) happening in real-time \citep{zacks_event_2007,sargent_event_2013,jeunehomme_event_2020,kurby_segmentation_2008,reynolds_computational_2007,gold_effects_2017,oetringer_2025_neural,baldassano_discovering_2017,de_bruin_shared_2023,Shin2020,de_soares_top-down_2023,zacks_prediction_2011,sava-segal_individual_2023,laing_event_2025}.
Behavioural and neural data indicate that event segmentation occurs automatically, where individuals spontaneously break continuous input into hierarchically organised parts and sub-parts, aiding in the comprehension, memory retention, and subsequent decision-making processes regarding the observed events \citep{kurby_segmentation_2008,zacks_event_2007,baldassano_discovering_2017,Baldassano2018,geerligs_partially_2022,oetringer_2025_neural}. 

Although alternative segmentation theories exist, Event Segmentation Theory (EST) specifically suggests that the brain constantly generates predictions about what will happen next based on current sensory input and past experiences \citep{zacks_event_2007}.
When the actual input deviates significantly from these predictions, the brain detects an event boundary and updates its internal models.
This process of segmentation structures various cognitive functions by aligning them with respect to these event boundaries.
Memory encoding, for instance, is thought to be triggered at event transitions \citep{ben-yakov_hippocampal_2018,Shin2020,sargent_event_2013,gold_effects_2017}.
Attention is also thought to be modulated as the brain reallocates resources to process new or unexpected information at the start of an event \citep{de_soares_top-down_2023,gozukaraMultiScaleAntiCorrelatedNeural2026,pradhanEventSegmentationEvent2022,swallowEventBoundariesPerception2009}.
By organizing perception, memory, attention, and prediction around these event structures, EST offers an explanation of how different cognitive functions interact and support each other in real time.

The neural study of event segmentation identifies `neural states' as a potential mechanism underlying event segmentation \citep{baldassano_discovering_2017, geerligs_partially_2022, oetringer_2025_neural, Baldassano2018}
In this context, neural states are defined as neural activity patterns that are stable for a given time period before they transition into different, but again temporarily stable, neural activity patterns.
These neural states are found to coincide with the temporal dynamics of relevant stimuli features \citep{oetringer_2025_neural}, are affected by attentional processes \citep{de_bruin_shared_2023, de_soares_top-down_2023}, are associated with memory \citep{baldassano_discovering_2017} and prediction mechanisms \citep{lee_anticipation_2021}, and are spatio-temporally nested across cortical scales \citep{geerligs_partially_2022,gozukaraMultiScaleAntiCorrelatedNeural2026}.
However, while ES describes the cognitive `what' and `why', it lacks a detailed description of the neural `how'.

The definition of a neural state studied in the context of event segmentation overlaps with the definition of an extensively studied neural phenomenon termed `metastability'.
Both neural states and metastable states involve quasi-stable states that are characterised by distinct spatio-temporal patterns of brain activity. 
Metastability is specifically defined as various circuits, regions, or networks working in temporary coordination which form these quasi-stable states  \citep{tognoli_metastable_2014,alderson_metastable_2020,la_camera_cortical_2019,borisyuk_metastable_2007,kurikawa_transitions_2021,brinkman_metastable_2022,mora-sanchez_scale-free_2019,roberts_metastable_2019,borisyuk_metastable_2007,beim_graben_metastable_2019}.
The metastability literature studies how cortical regions maintain their individual autonomy and specialised functions (segregation, modularity) while coupling and coordinating globally for multiple functions (integration) \citep{wolff_intrinsic_2022,tognoli_metastable_2014,hancock_metastability_2025,alderson_metastable_2020}.
Metastability in continuous neural and behavioural recordings has been shown in multiple species and tasks, and it has been suggested that metastable dynamics underlie the real-time coordination necessary for the brain's dynamic cognitive, behavioural, and social functions \citep{alderson_metastable_2020,benozzo_slower_2021,la_camera_cortical_2019,mazzucato_expectation-induced_2019,recanatesi_metastable_2022}.
This body of literature focuses on the dynamic transition of metastable neural activity (MNA) states and aims to identify the characteristics of dynamical systems that would operate in similar metastable manners.

ES is a primarily cognitive framework with its roots embedded in behavioural investigations; while MNA is primarily computational, describing and modelling neural dynamics under various conditions.
The aim of this review is to demonstrate that the cognitive framework offered by ES and empirical findings about brain dynamics demonstrated by the metastability literature offer a complementary perspective on naturalistic real-time brain dynamics.
We will integrate the contemporary state of the ES and metastability literatures across various domains of cognitive neuroscience, including perception, decision-making, action, predictive coding, and memory.
We will also offer a set of core principles regarding the properties of neural states and how they operate across these different domains.

The remaining sections will first (1) discuss the working definition of neural states and their presence across spatio-temporal scales, second (2) discuss how they relate to the perception-action loop, memory, and predictive processing, and finally third (3) summarize important core principles of neural states and describe empirical questions that can be addressed using this perspective.

\section{Neural States}
In this paper, we will use the term neural states to refer to the temporarily stable neural activity patterns that have been observed in event segmentation literature as well as to metastable states that have been described in the MNA literature.
To understand the term ``neural state", imagine the activity levels of a set of N units (neurons, voxels or electrodes) as points in an N-dimensional space, where each axis corresponds to the activity level of a given unit.
This is called a \textbf{state-space}, and a neural state is a point in this space.
Instead of moving continuously through this space, neural activity jumps between distinct points or patterns in this state-space, called \textbf{attractors}.
These attractors are patterns or states toward which neural activity gravitates.
Metastability mainly refers to the fact that neural systems hover near attractors without fully stabilizing, meaning that both the exact points they hover around and the trajectory they take can dynamically change  \citep[as opposed to terms like multistability which refers to multiple but definite attractors and trajectories;][]{recanatesi_metastable_2022,wimmer_bump_2014,deco_ongoing_2012,varela_brainweb_2001,kurikawa_transitions_2021,}.
Metastability is used to reason on and model flexible switches between neural states in an attempt to understand how this phenomenon enables dynamic mental functions like learning, cognition, or action \citep{hulsey_decision-making_2023,siems_rhythmic_2023,ponce-alvarez_dynamics_2012,recanatesi_metastable_2022,benozzo_slower_2021,la_camera_cortical_2019}.


The robustness of neural states is evidenced by their detection via diverse analytical approaches.
Since as early as 1995, activity across multiple single cell recordings have been described as a sequence of neural states, using hidden Markov models  \citep[HMMs;][]{abeles_cortical_1995,kemereDetectingNeuralStateTransitions2008,jonesNaturalStimuliEvoke2007,escolaHiddenMarkovModels2011, baldassano_discovering_2017, vidaurreSpontaneousCorticalActivity2018}.
More recently, specialized methods are being developed to describe neural states, such as Greedy State Boundary Search (GSBS) which are being used alongside HMMs in neuroimaging data analysis \citep{geerligs_detecting_2021,geerligs_partially_2022}.
There are also numerous other approaches such as control theory, dynamical systems theory, chaos theory, and more that can be used to describe state-like activity in contemporary neuroscience \citep{johnItsTimeLinking2022,rabinovichDynamicalPrinciplesNeuroscience2006,guControllabilityStructuralBrain2015,stamNonlinearDynamicalAnalysis2005,kornThereChaosBrain2003,breakspearDynamicModelsLargescale2017a}.
Depending on the choice of recording units and analysis, a neural state can describe a localized population of neurons, the global activity of the whole brain, and anything in between.
Indeed, states have been studied across various temporal and spatial scales, from sub-second states in local populations, to global whole-brain states that can last for minutes or even days \citep{abeles_cortical_1995,oetringer_2025_neural,brady_differential_2017,marques_internal_2020}. 

Below, we summarize a range of empirical findings that identify neural states at a range of temporal (milliseconds, seconds, minutes...) and spatial (cells, areas, networks...) scales.

\subsection{Temporal and Spatial Scales of Neural States}
According to ES, events are nested in a temporal hierarchy of longer and more abstract cognitive events, similar to the `spatially-semantically' nested hierarchies of perceptual representations \citep{zacks_event_2007, zacksPerceivingRememberingCommunicating2001,tverskyStructureExperience2008,kurby_segmentation_2008, radvanskyEventCognition2014,hardShapeAction2011,dubrowInfluenceContextBoundaries2013}.
An event at an arbitrary level of abstraction, such as `Preparing breakfast' can be broken down into sub-events such as `Brewing coffee', `Cooking eggs', `Cutting bread', `Eating', and so on.
These sub events can further be segmented down in steps until the smallest behaviourally relevant events are reached (e.g, from more to less abstract; boil water for coffee, fill boiler with water, turn the tap on, move hand down...).
This mirrors the `spatially-semantically' nested hierarchies of object definitions, where a house can be broken down into sub-parts such as walls, doors, floor, ceiling, and so on; which can themselves be segmented down into smaller units like bricks and planks \citep{zacksEventPerceptionMind2007,zacksEventStructurePerception2001a}.
Neural investigation of ES supports the idea that there is an analogous neural structure to the cognitive one, with successively higher order cortical areas showing longer state durations, reflecting their cognitive counterparts \citep{geerligs_partially_2022,baldassano_discovering_2017,honey_slow_2012, Baldassano2018, oetringer_2025_neural}.

Descriptions of nested hierarchies are also prevalent in the MNA literature.
For instance, \citet{varela_brainweb_2001} discuss metastable dynamics at three different temporal scales: the scale of cellular rhythms (10–100 ms), the scale of large scale integration relevant for the transitions between dynamic core configurations (100–300 ms), and the scale of long-range integration ($>$1 s).
Since then, metastable states have been observed across a much wider temporal range.
Electrophysiological studies targeting different spatio-temporal levels show evidence of sub-second neural states in local neural populations \citep{abeles_cortical_1995,maboudi_uncovering_2018}, and in whole-brain EEG microstates \citep{von_wegner_eeg_2021}.
States in the range of seconds to minutes associated with various cognitive and behavioural phenomena are also reported, for example in relation to working memory and decision making \citep{ponce-alvarez_dynamics_2012,taghia_uncovering_2018}.
On the same timescale, fMRI resting state studies report discrete sets of coordinated functional networks and characterize global brain activity as the exploration of a series of states associated with different functional networks \citep{song_large-scale_2023,bolt_parsimonious_2022,thompson_quasi-periodic_2014,karahanoglu_transient_2015}.
States lasting minutes to hours have also been observed.
For example, \citet{marques_internal_2020} describe metastable states corresponding to explorative and exploitative states in foraging behaviour.
Lastly, metastable states might even occur on the timescale of days, as shown for example by the relationship between functional network connectivity patterns and with mood changes in bipolar disorder \citep{brady_differential_2017}.

The examples in the previous paragraph already suggest that in addition to temporal scales, neural states also exist across different spatial scales.
Generally, shorter states tend to occur at more local spatial scales, such as the sub-second states that occur in local neural populations \citep{abeles_cortical_1995,benozzo_slower_2021,ponce-alvarez_dynamics_2012,recanatesi_metastable_2022}, while longer states, for example those on the timescale of multiple seconds to minutes, involve whole brain networks \citep{lee_emergence_2017,de_bruin_shared_2023,yousefi_propagating_2021,thompson_quasi-periodic_2014,gozukaraMultiScaleAntiCorrelatedNeural2026,song_hierarchical_2023,song_large-scale_2023,yamashita_brain_2021}.
Note though that this is not necessarily always the case, with observations, for example, of whole brain EEG microstates at the sub-second scale \citep{von_wegner_eeg_2021,tait_meg_2022,michalopoulos_combining_2015,michel_eeg_2018}.
The idea of spatial nestedness of neural states is also in line with our current understanding of interactions between spatial scales of brain function, where local microcircuits interact to create mesoscale networks of brain regions, which in turn influence macroscale systems involving large networks across the brain \citep{varela_brainweb_2001}.


There are some suggestions in the literature about the neural mechanisms that underlie this spatial and temporal nestedness of neural states. 
At the shortest timescales, cortical neural populations simultaneously alternate between periods of vigorous spiking activity (often termed on/up states) and periods of relative silence (off/down states).
First characterised during slow-wave sleep as cortical slow oscillations \citep{steriadeThalamocorticalOscillationsSleeping1993,steriadeNeuronalPlasticityThalamocortical2003}, these fluctuations are now known to occur during wakefulness as well, as rapid, local modulations of population excitability \citep{la_camera_cortical_2019,mazzucatoDynamicsMultistableStates2015}.
In primate visual cortex, for example, ensemble activity in V4 spontaneously fluctuate between on and off phases synchronously across cortical layer, and the momentary phase of this alternation predicts behavioural performance \citep{engel_selective_2016}.

Within on states, cortical activity is not homogenous but itself also unfolds as sequential 'packets' of spiking activity lasting from 50 to 200 ms, both spontaneously and in response to stimuli \citep{luczakPacketbasedCommunicationCortex2015}.
These packets carry information about stimulus identity through variation in the timing and number of spikes within a sequential structure.
In this sense, on states themselves have internal temporal structure that can encode information, forming perhaps the finest-grained units in the neural state hierarchy.

Brain oscillations of different frequencies provide the temporal scaffolding for these dynamics.
Oscillatory cycles create alternating windows of high and low neural excitability, and when two populations are phase-aligned in their oscillatory activity, the excitatory windows coincide, enabling effective communication between them \citep{fries_rhythms_2015,engelDynamicPredictionsOscillations2001}.
This principle, formalised as `communication through coherence' by \citet{fries_rhythms_2015}, suggests that stable phase relationships between oscillators are the driving mechanism behind selective functional information flow between neural populations.
When such phase relationships are sustained, they give rise to longer metastable states on the order of hundreds of milliseconds to seconds, as populations that are coherently coupled maintain coordinated activity patterns \citep{roberts_metastable_2019,varela_brainweb_2001}.
Computational modelling confirms this, as stable phase coupling between oscillatory signals at different frequencies can produce slower metastable dynamics in the range of seconds \citep{roberts_metastable_2019}.

These phase-coupled dynamics also have a spatial dimension that is increasingly recognised as fundamental to the emergence of brain states.
Computational modelling suggests that when oscillatory activity propagates across spatially extended cortical tissue, it does so with a rich repertoire of wave patterns, including travelling waves, spiral waves, sources, and sinks \citep{roberts_metastable_2019}.
These wave patterns are themselves metastable, as in the system visits multiple spatio-temporal configurations in sequence, with transitions between them corresponding to nonlinear instabilities in the underlying phase flows, such as dissolution, emergence, or collision of wave sources and sinks.
This means that the metastable neural states described throughout this review can be understood as transient wave patterns whose stability depends on the maintenance of coherent phase relationships across cortical regions \citep{roberts_metastable_2019}.

Empirical evidence supports this wave-based picture across multiple spatial and temporal scales.
In fMRI, spiral-like rotational wave patterns have been shown to be widespread during both rest and cognitive task states, with their properties (such as rotational direction and cortical location) varying systematically across task and predicting cognitive processing \citep{xu_interacting_2023}.
Multiple interacting spirals coordinate the correlated activations and deactivations of distributed functional regions enabling flexible reconfigurations of activity flow between bottom-up and top-down directions.
At slower timescales, infra-slow waves (\textless 0.1 Hz) propagate across the entire brain in parallel, travelling systematically from unimodal sensory regions towards transmodal association cortex along the brains principal functional connectivity gradient \citep{marguliesSituatingDefaultmodeNetwork2016,raut_global_2021,thompsonQuasiperiodicPatternsQPP2014}.
These global waves synchronise the brains functional systems with fluctuating arousal, such that different temporal phases of a 10 to 40 second spatio-temporal cycle correspond to different spatial patterns of enhanced excitability.
Recent work suggests that the repertoire of possible wave patterns is partly constrained by the physical geometry of the brain.
\citet{pangGeometricConstraintsHuman2023} demonstrated that cortical activity can be parsimoniously reconstructed using geometric eigenmodes (resonant modes) derived from the shape of the cortical surface.
Building on these, \citet{fosterBrainStatesWavelike2024} propose that brain states could be reconceptualised as 'wave-like motifs', i.e. recurring patterns of propagating activity that can be decomposed into a small number of spatio-temporal propagation modes.

Together, these findings suggest that the metastable neural states at the core of this review have an inherently spatio-temporal character.
The metastable dynamics we emphasise (where stable states are punctuated by rapid transitions), arise from the propagation of waves across cortical geometry, with transitions corresponding to instabilities in wave patterns.

\subsection{Neural State Transitions} \label{Neural State Transitions}

So, what determines when one stable state transitions to the next?
The ES and MNA literatures offer complementary perspectives on this question.

MNA literature suggests that state boundaries are driven by dynamically changing functional connectivity networks.
Around state boundaries, the neural communication range is maximised, allowing brain areas that are far away from one another to influence each other \citep{shineDynamicsFunctionalBrain2016, tognoli_metastable_2014, decoRethinkingSegregationIntegration2015,hancock_metastability_2025}.
Under this perspective, neural state boundaries are brief temporal windows where the information integration capacity and adaptability of the network is at its highest \citep{shineDynamicsFunctionalBrain2016,alderson_metastable_2020,spornsNetworkAttributesSegregation2013}.
Within a neural state, the functional connectivity of a given neural population would be more constrained.

From the ES side, event boundaries have been linked to changes in context or task demands.
One proposal is that event boundaries occur when a new context (or event type) is detected \citep{shin_structuring_2021} or when there are changes in tasks \citep{clewett_transcending_2019,wang_switching_2022}.
In naturalistic settings such as movie watching or story listening, these translate as narrative events and/or continuity-providing stimuli features \citep{oetringer_2025_neural, geerligs_partially_2022, baldassano_discovering_2017}.
This may facilitate freeing up space in working memory to focus on the most relevant information for our current goal \citep{shim_generating_2024}.
A closely related account holds that event boundaries are driven by prediction errors \citep{zacks_prediction_2011}, where cognitive (and neural) states can be thought of as predictive schemas maintained until bottom-up or top-down influences decrease the predictive accuracy of the schema, at which point a new event model is initiated.
There is some behavioural and neural evidence for this view: decreases in stimulus feature predictability have been associated with event and neural state boundaries \citep{zacks_prediction_2011, pettijohnNarrativeEventBoundaries2016,eisenbergDynamicPredictionPerception2018a}, and predictability of within-event items is shown to be higher than between-event items, indicating an underlying schema-like structure \citep{Shin2020,Baldassano2018,zacks_prediction_2011,franklinStructuredEventMemory,pettijohnNarrativeEventBoundaries2016}.
Context-change has been proposed as a distinct alternative account for driving event boundaries \cite{clewett_transcending_2019,shin_structuring_2021,wang_switching_2022,heusserPerceptualBoundariesCause2018,barorRoleContextContinuity2026}, however these two account (predictability and context-change) might point towards the same underlying structure, as we discuss in the section on predictive processing below.

Combining these two perspectives, ES and MNA together suggest a hierarchy of stable states, persisting until a reconfiguration of the network is triggered, which requires a greater amount of information integration.
This implies that neural state boundaries correspond to moments of high integration and communication, while the stable event periods correspond to segregated and modular processing.

\section{Neural States as Units of Cognition}

After presenting a more concrete idea on what we mean by neural states, we can now look at naturalistic continuous brain activity through the lens of spatio-temporal metastable neural states in order to form a coherent view on how different aspects of cognition integrate during continuous naturalistic behaviour.
In this section we will investigate how naturalistic perception, decision making, action, prediction, and memory are orchestrated via neural states.

\subsection{Perception, Decision Making, and Action}

It is well known that brain activity in response to perceptual input is composed of increasingly abstract neural representations across the cortical hierarchy \citep{gucluDeepNeuralNetworks2015,dicarloHowDoesBrain2012,UsingGoaldrivenDeep,fellemanDistributedHierarchicalProcessing1991}.
While these perceptual representations are typically investigated in the spatial dimension (i.e. by investigating which brain regions encode a specific stimulus representation), neural states highlight the temporal nature of these perceptual representations \citep{geerligs_partially_2022,geerligs_detecting_2021,baldassano_discovering_2017,oetringer_2025_neural,Baldassano2018}.
A temporal neural state hierarchy analogous to the spatial one has been observed in a range of studies and shows the expected gradient of short neural states in primary sensory regions, and long states in higher order regions such as the frontal and temporal cortices \citep{geerligs_partially_2022}.

\subsubsection{Perception}

To draw a more concrete image of how neural states underlie dynamic perception, let's traverse the nested states hierarchy in the context of vision.
We can start with saccade-coupled neural states in theta frequency oscillations in primary visual areas.
At each fixation of the eyes, there is a synchronization of spikes in area V1 in monkeys, suggesting that local field potential modulations are locked to saccade onsets, similar to oscillation driven on-off states \citep{ito_saccade-related_2011}.
Detailing this view, \citet{wutz_temporal_2016} show that in humans, V1 neural temporal integration windows align with saccades, potentially forming the smallest units of visual neural states.
They suggest that this alignment serves as an organising principle for the temporal processing of continuous sensory input in vision \citep{wutz_temporal_2016}.
Further up the ventral and dorsal visual streams, we find neural states that represent the temporal stability of increasingly complex visual features in the environment.
For example, \citet{oetringer_2025_neural} show that changes in the scene locations in a movie are accompanied by neural state changes in the ventral stream areas such as PPA and RSC.
Furthermore they show that changes in simpler visual features coincide with changes in early visual cortex states, providing more evidence for the hierarchical relationship between representations at different abstraction levels \citep{oetringer_2025_neural}.
Even longer neural states have been linked to even more abstract events, both in higher order areas like the angular gyrus, medial prefrontal cortex, anterior insula and anterior cingulate and whole brain networks like the default mode network \citep{baldassano_discovering_2017,de_bruin_shared_2023,geerligs_partially_2022}.

The notion of scanpaths by \citet{noton_scanpaths_1971} is a good example of a specific sequence nested within a temporally stable representation of a higher abstraction level. 
These scanpaths contain sequences of saccades that are present during object recognition and are later observed again during recognition or recall \citep{noton_scanpaths_1971,kelley_repetitive_2021,johansson_eye-movement_2022,wynn_eye_2024}. 
The limited size of the high-resolution fovea favours sampling information from the most informative features of the object, (like the corners of a chair, eyes, nose, and mouth of a face, etc.), and their relative position  \citep[path of the scan;][]{yarbusEyeMovementsVision1967,hendersonHumanGazeControl2003,petersonLookingJustEyes2012,hayhoeEyeMovementsNatural2005}.
These individual fixations result in individual neural states \citep{ito_saccade-related_2011,wutz_temporal_2016}, which are themselves nested within a more stable object representation state.
The stable object representation may be speculated to acts as an anchor that holds these neural state sequences together.
Alongside perceiving an eye, nose, and a mouth, we also perceive a face, perhaps thanks to higher-order visual areas which maintain stable states that span the duration of the entire scanpath.

\subsubsection{Decision Making and Action}

Although so far we have mainly related neural states to elements of perception, there is good evidence to suggest that decision making and action also occur within this neural state hierarchy.
In fact, as demonstrated with the saccade example above, we prefer not to view perception, decision making, and action as separate processes but rather as different perspectives on the same process \citep{hommel_theory_2019}.
Each action we take, results in updates of our perceptual and cognitive models, which in turn drives our decision-making and action processes again.
The scanpath example above reflects the inherent connection between perception, cognition, and action.
Saccadic pattern sequences are repeated during visual object recognition, as well as recall, where we could imagine they may not be necessary. 
From this perspective, decision making can be seen as the process that determines which specific state pattern will follow the current one and when.

The MNA literature suggests that one of the mechanisms behind state transitions could be a winner-less competition between relevant modes \citep{rabinovichDynamicalEncodingNetworks2001,rabinovichTransientCognitiveDynamics2008,rabinovichRobustTransientDynamics2011,afraimovichHeteroclinicContoursNeural2004,meyer-ortmannsHeteroclinicNetworksBrain2023}.
This view is in line with the affordance competition hypothesis put forward by \citet{cisek_cortical_2007}, which describes action selection as competing neural codes which are biased by top-down and bottom-up input from other areas.
We can think of competing neural states as competing activity state attractors, with biases coming from top down and bottom up influences determining the ‘winning’ neural representation by altering the landscape.
Under this view, constant interactions between internal dynamics of and external inputs to a neural population/network would give rise to temporarily stable states which are sustained until another mode `wins' the competition and changes the neural state, which shapes the decision that is made on the behavioural level.
In line with this view, neural state durations have been linked to the outcomes of decision making processes in both human and animal work \citep{benozzo_slower_2021,taghia_uncovering_2018}.
For example, during distance discrimination tasks in mice, longer neural states tend to precede errors and difficult trials lead to delays in state transitions compared to shorter easy trails \citep{benozzo_slower_2021}.
These studies suggest decision making is underpinned by local neural state transitions. 

Local neural states that drive specific decisions are strongly shaped by minutes-long whole-brain states associated with performance levels and exploitative versus explorative behaviour \citep{marques_internal_2020,hulsey_decision-making_2023,yamashita_variable_2021,song_hierarchical_2023}.
In mice, transitions between three distinct minute-long behavioural states have been observed, where optimal levels of task performance were linked to a specific state \citep{hulsey_decision-making_2023}.
In humans, distinct whole brain neural states have been described during naturalistic conditions like movie watching or reading, as well as in resting state \citep{bolt_parsimonious_2022,song_large-scale_2023,gozukaraMultiScaleAntiCorrelatedNeural2026,yangDefaultNetworkDominates2023,brandmanSurprisingRoleDefault2021}.
These states have been linked to arousal \citep{raut_global_2021}, show associations with behaviour and performance, and are modulated by individual differences, for example in the context of attention deficit disorder \citep{yamashita_variable_2021}, ageing \citep{lugtmeijerTemporalDedifferentiationNeural2025,stawarczykAgingEncodingChanges2020,reaghAgingAltersNeural2020,kurbyAgeDifferencesPerception2011}, or general memory performance \citep{gozukaraMemoryHippocampalResponses2026a}.
These findings imply that fluctuations in whole brain states shape the transition probabilities of local metastable states that govern decision making and motor processes.

\subsubsection{Attention}

A growing body of evidence suggests that attention also samples the environment in discrete temporal windows governed by neural oscillations, primarily in the theta (3-8 Hz) and alpha (8-12 Hz) bands \citep{vanrullenPerceptualCycles2016, landauAttentionSamplesStimuli2012, helfrichNeuralMechanismsSustained2018}.
\citet{vanrullenPerceptualCycles2016} distinguishes between two coexisting perceptual rhythms in vision: an alpha rhythm ($\sim$10 Hz) associated with sensory excitability windows that gate what can be perceived, and a theta rhythm ($\sim$7 Hz) associated with attentional selection of what is perceived.
\citet{fiebelkornRhythmicTheoryAttention2019} propose that theta oscillations in the attention network also coordinate the alternation between perceptual sampling and motor preparation.
During one phase of the theta cycle, functional connections between higher-order regions and sensory areas are enhanced, promoting sampling at the currently attended location.
During the opposite phase, connections to motor regions are favoured, facilitating exploratory movements such as saccades.
This rhythmic structure connects naturally to the saccade-coupled states described at the beginning of this section.
Each saccade resets the phase of theta oscillations, including in the hippocampus, effectively initiating a new processing cycle at each fixation \citep{ito_saccade-related_2011}.
In this sense, the oscillation-driven on and off states described in the neural mechanisms section are the temporal substrate through which attention actively structures the perception-action loop.

Together, these findings suggest that the hierarchical organisation of neural states is an integral part of the spatio-temporal perception-action loop.
Perception, decision making, action, and attention all operate through the same nested state dynamics, where longer-duration states in higher-order regions constrain the range of possible states in faster-operating areas, and faster ones lead to constraints on the slower ones in return.
Crucially, this constraining influence is not limited to concurrent processing.
The same mechanism through which a current whole-brain state restricts the landscape of local neural states could also operate across time, with past states shaping future ones through prediction, and with retrieved states from memory re-entering the hierarchy to influence ongoing processing \citep{de_soares_top-down_2023, engel_selective_2016, papale_modulatory_2023}.
In the following two sections, we develop this idea for prediction and memory in turn.

\subsection{Predictive Processing}

The idea that the brain is fundamentally oriented towards the future has become one of the most influential frameworks in cognitive neuroscience \citep{clark_whatever_2013, friston_free-energy_2010}.
Rather than being passive receivers of information, our brains are thought to continuously generate and revise predictions about incoming sensory data.
However, terms such as \textit{predictive coding}, \textit{predictive processing}, \textit{prediction error}, \textit{expectation}, and \textit{anticipation} are frequently used interchangeably in the literature, creating a landscape where genuine theoretical disagreements are difficult to distinguish from differences in terminology.
For instance, \textit{predictive coding} is often used as an umbrella term, but it is technically a specific algorithm making commitments about how prediction errors are computed and passed between cortical layers.
Broader claims such as that the brain minimises prediction error, performs probabilistic inference, or uses generative models are logically independent of that algorithm and have empirical support in their own right \citep{sprevakPredictiveCodingIntroduction2024}.
Similarly, \textit{prediction} can refer to memory-based expectations about future events, active neural anticipation that pre-activates sensory areas, longer-timescale prospection, or an umbrella term for all future-oriented processing \citep{bubicPredictionCognitionBrain2010}; and predictive information in the brain is not exclusively top-down but is also embedded in bottom-up processing through learned statistical regularities \citep{teufelFormsPredictionNervous2020}.
We raise these distinctions because our proposal below draws on the broader framework, instead of the narrow algorithm.
With these distinctions in mind, we can now be more precise about how metastable neural states relate to prediction.

We propose that metastable neural states are a reflection of interaction between incoming signals and generative internal models encoding an internally generated expectation of said signals \citep{friston_free-energy_2010, clark_whatever_2013}.
In the context of the spatio-temporal hierarchy described in previous sections, this means that a neural state at a given hierarchical level is used to generate expectations appropriate for the timescale at which that level operates.
Higher-order states in regions such as the angular gyrus or medial prefrontal cortex would be the basis for generating expectations about abstract, slowly changing features (e.g., the current narrative context or social situation), while states in early sensory areas could be used to generate expectations about rapidly changing low-level features  \citep[e.g., the trajectory of a moving object or the spectral content of speech;][]{caucheteux_evidence_2023, lee_anticipation_2021}.
These expectations combine both top-down contextual predictions and bottom-up statistical regularities embodied in the activation patterns themselves \citep{teufelFormsPredictionNervous2020, delangeHowExpectationsShape2018}.
Importantly, these generative models are not based upon static templates.
During the stable period of a neural state, the model is continuously refined as the circuit accumulates information over time.
This accumulation is hierarchically organised such that early sensory circuits integrate over tens of milliseconds, intermediate areas over seconds, and higher-order regions over tens of seconds or longer \citep{hassonHierarchicalProcessMemory2015, honey_slow_2012}.

If neural states are indeed the representational reflections of generative models, we would expect this to have consequences for prediction accuracy, state transitions, and state timing.
Here we look at each of these in turn.

Regarding prediction accuracy, if a neural state is a reflection of a coherent generative model, we should expect not only that predictions are more accurate within stable states than at their boundaries \citep{zacks_prediction_2011, eisenbergDynamicPredictionPerception2018a}, but also that neural states themselves carry signatures of active generation.
The phenomenon of anticipatory reinstatement provides evidence for this.
When people have strong expectations about upcoming events based on prior experience, neural states associated with those events are reinstated before the actual stimulus appears \citep{lee_anticipation_2021}.
This suggests that neural states are not passive reflections of current input but are actively produced by the underlying model, shaped by accumulated past context.
At first glance, these two findings may appear contradictory.
If boundaries are moments of prediction failure, how can they also be anticipated?
The resolution lies in the hierarchical nature of the generative model.
Prediction of 'what will happen within a context' and prediction 'that a context will change' are not the same computation and likely operate at different levels of the hierarchy.
A higher-order model that has learned the temporal structure of an experience can successfully predict an upcoming transition to drive anticipatory reinstatement, while content-level predictions about specific sensory input are nonetheless disrupted as one generative model gives way to the next.
This can be seen as a reflection of the generative underlying nature of neural states where the model is actively producing expected patterns of activity ahead of the incoming input, shaped by what has been accumulated from past experience.

Regarding state transitions, if neural states reflect underlying generative models, then boundaries should correspond to moments when the model must be substantially updated.
As mentioned in sections above, both prediction error and context change have been proposed as drivers of event boundaries \citep{zacks_prediction_2011,clewett_transcending_2019,shin_structuring_2021,wang_switching_2022,eisenbergDynamicPredictionPerception2018a}.
The generative model perspective suggests these are not competing accounts.
A context change, even a predictable one, requires a reconfiguration of the generative model, because the expected patterns of activity under the new context are fundamentally different from those under the old context.
This is better captured by Bayesian surprise (how much the model's beliefs must change) than by surprisal (how unlikely a specific observation is).
Indeed, Bayesian surprise has been shown to be a better predictor of event segmentation during language listening \citep{kumarBayesianSurprisePredicts2023}.
This also aligns with the MNA literature's characterisation of state boundaries as windows of heightened information integration and connectivity reconfiguration as updating a generative model requires the kind of broad neural communication that is observed at boundaries \citep{shineDynamicsFunctionalBrain2016,tognoli_metastable_2014, hancock_metastability_2025,decoRethinkingSegregationIntegration2015}.

Finally, if neural states are a reflection of underlying generative models, their temporal dynamics should depend on the predictability of the environment.
States should persist longer when the environment is consistent with the current model and transition more rapidly when it is not.
Indeed, for instance, \citet{lee_anticipation_2021} show that neural state boundaries shift earlier in time with increasing familiarity, and that this effect is stronger in higher-order multimodal areas than in sensory regions, consistent with a hierarchical generative model framework.
Beyond timing, recent work leveraging large language model text embeddings has demonstrated that neural representations explicitly encode predictive context at multiple timescales \citep{caucheteux_evidence_2023}, supporting the view that what neural states constitute models that blend past context and future expectation.

Overall, these studies suggest that metastable neural states are not passive representations of incoming information but are reflections of active generative models that constrain perception, guide action, and shape memory.
This view also reveals an intimate link between prediction and memory.
The generative model that a neural state reflects is built from accumulated past information, what \citet{hassonHierarchicalProcessMemory2015} term 'process memory'.
Based on this aggregated literature, we could speculate that the accumulated context that enables prediction could be considered a form of memory, and memory encoding at state boundaries could be considered to be processes of consolidation of the current generative model, updating relevant old ones, and finally initialising the next.
We develop this connection further in the following section.

\subsection{Memory}

Crucially, the brain does not only process states for immediate action and prediction but also stores important states and their sequences to guide future behaviour.
Memory is often associated with event segmentation and has been extensively studied in the context of neural states.
Memory processes are intertwined with the hierarchical organisation of neural states, working through the same metastable dynamics that underlie perception and action.

A useful conceptual framework for understanding this relationship is the Hierarchical Process Memory (HPM) model proposed by \citet{hassonHierarchicalProcessMemory2015}.
HPM argues that memory is not restricted to a few localised stores that are functionally separate from processing.
Instead, virtually all cortical circuits accumulate information over time, and the timescale of this accumulation varies hierarchically; from tens of milliseconds in early sensory areas, to seconds in intermediate areas, to minutes in higher-order regions such as the angular gyrus, medial prefrontal cortex, and retrosplenial cortex.
In this view, memory is intrinsic to information processing itself.
Past information continuously shapes how incoming information is processed at every level of the cortical hierarchy.

The HPM framework maps naturally onto the perspective on metastable neural states that we developed in this review.
The hierarchical timescales of process memory correspond to the nested temporal hierarchy of neural states we have described throughout this review (brief states in sensory cortex, longer states in association areas, and the longest states in prefrontal and default-mode regions).
What HPM adds is the explicit claim that each of these levels constitutes a form of memory, continuously integrating past context with present input.
What the work we discussed on neural states and metastability adds to HPM is an account of what happens at the transitions between these accumulated windows.
At state boundaries, functional connectivity reconfigures and the hippocampus can exploit these moments to bind distributed representations into episodic memory traces.
In addition, it has been well established in event segmentation literature that an event boundary results in a clearing of working memory, which we can conceptualize as a reset of this continuous integration process \citep{gulerDiscreteMemoriesContinuous2024,ongchocoDidThatJust2019,ezzyatWhatConstitutesEpisode2011a,clewettEbbFlowExperience2017}. 

Below, we describe how neural state dynamics and memory interact through at least three interconnected mechanisms: (1) boundary-triggered encoding, (2) pattern-based storage and replay, and (3) state-dependent retrieval. 

\subsubsection{Boundary Triggered Memory Encoding}

Numerous studies have shown that transition points from one neural state (or event) to the next are associated with storage and recall of memory.
It has been well documented that event boundaries (such as shot transitions in naturalistic video clips), are accompanied by an increase in hippocampal activity \citep{baldassano_discovering_2017,ben-yakov_hippocampal_2018,hahamy_human_2022,shin_structuring_2021,swallow_changes_2011}.
Neural state boundaries in areas such as the precuneus and angular gyrus; as well as whole-brain neural state boundaries have also been linked to an increase in hippocampal activity and better memory \citep{ben-yakov_hippocampal_2018,baldassano_discovering_2017,gozukaraMemoryHippocampalResponses2026a}.
Furthermore, increased hippocampal activity at these boundaries is associated with better memory performance, supporting the idea that the hippocampal activity indeed reflects the encoding of events \citep{ben-yakovConstructingRealisticEngrams2011a,ben-yakovHippocampalImmediatePoststimulus2013a,dubrowTemporalBindingEvents2016a,baldassano_discovering_2017,reaghAgingAltersNeural2020}.

In the section on neural state transitions, we have proposed that state boundaries are moments of maximal information integration, when functional connectivity is reconfigured and new predictive models are formed. 
Within the HPM framework, these boundaries can be understood as moments when the accumulated process memory of one state is finalised and a new accumulation begins.
The hippocampus appears to exploit these transition points to bind together the elements of an experience into a coherent memory trace.
This occurs throughout the nested hierarchy: during a state boundary in higher-order cortical areas, local populations across multiple other regions are also often reconfiguring their activity patterns simultaneously \citep{geerligs_partially_2022,gozukaraMultiScaleAntiCorrelatedNeural2026}.
The hippocampus, receiving convergent input from these areas, can capture this distributed pattern to be crystallized.
This suggests that the hippocampus is encoding the structured units defined by cortical state dynamics.

Individual differences in event segmentation ability support this view.
People who naturally segment experiences into clearer, more distinct events show better memory performance \citep{kurby_segmentation_2008,sargent_event_2013}.
When external cues help people segment more effectively, their memory improves correspondingly \citep{gold_effects_2017}.
This makes sense if memory encoding is locked to state boundaries as clearer boundaries would create more distinct memory traces.

\subsubsection{Pattern Based Storage and Replay}

In the previous section, we argued that neural states are reflections of underlying generative models that accumulate context and generate expectations.
Throughout this section, we describe how neural activity patterns associated with these states are stored, replayed, and reinstated during memory processes.
It is important to note that these observable activity patterns are the measurable signatures of the generative models described above.
When we observe pattern reinstatement during recall, what we are measuring is a generative model re-activating and producing the activity that characterises its operation.
In this sense, memory engrams are themselves models, and pattern reinstatement reflects their re-engagement.

During ongoing experience, populations of neurons settle into metastable configurations that persist for seconds or longer.
These same configurations are later reinstated during memory-related processes, suggesting that the underlying models themselves are preserved \citep{hassonHierarchicalProcessMemory2015}.
This aligns with the HPM perspective that the accumulated activity patterns within cortical circuits constitute both the processing and the memory.
We can find support for this starting from ongoing naturalistic activity and rest.
During movie-watching, it has been shown that information from the previous event is briefly reactivated right after an event boundary, providing more evidence for memory formation and consolidation at these moments \citep{hahamy_human_2022, silva_rapid_2019}.
At the same time, older but relevant state patterns also replay at event boundaries, which may help to piece together relevant parts of an ongoing experience and maintain context \citep{hahamy_human_2022}.
These kinds of neural state pattern replays are also seen during memory consolidation in rest.
When animals rest after behaviour, hippocampal sharp-wave ripples replay sequences of neural states that occurred during the task, but compressed in time \citep{maboudi_uncovering_2018}. 
A state that lasted seconds during behaviour might be replayed in just tens of milliseconds during a ripple, but the state patterns and sequence order are preserved.

Similar reinstatement occurs during retrieval too.
When people recall an event, cortical regions reinstate the neural state patterns that were active during the original experience \citep{Baldassano2018,chen_shared_2017}.
The fidelity of this reinstatement predicts retrieval accuracy, suggesting that successful remembering involves recreating the state patterns and sequences that originally represented the experience \citep{oedekovenReinstatementMemoryRepresentations2017,baldassano_discovering_2017,jeunehommeRepresentationalDynamicsMemories2022,stawarczykAgingEncodingChanges2020}.
Anticipatory reinstatement is particularly informative in this regard.
When people have strong predictions about upcoming events based on memory, neural states associated with those events appear before the actual stimulus.
Specifically, anticipatory reinstatement follows the spatio-temporal cortical hierarchy, where faster regions anticipate in shorter timescales and slower ones longer \citep{lee_anticipation_2021}.
This hints at memory retrieval operating across the same bidirectional hierarchy as the perception-action loop, with higher-order regions driving active recall through top-down reinstatement while sensory regions support recognition through bottom-up pattern matching.

\subsubsection{State Dependent Encoding and Retrieval}

Where in time a particular moment falls relative to state boundaries, affects how well that moment is remembered.
Moments near event boundaries are more likely to be encoded and later recalled than moments in the middle of stable states \citep{jeunehomme_event_2020, swallowEventBoundariesPerception2009}.
At the same time, event boundaries make it more difficult to access details from the preceding event, as demonstrated by the finding that within-event cued recall consistently exceeds across-boundary cued recall \citep{ezzyatWhatConstitutesEpisode2011a, radvanskyWalkingDoorwaysCauses2011, radvanskyWalkingDoorwaysCauses2006}, and that even actively maintained working memory contents are disrupted by event boundaries \citep{gulerDiscreteMemoriesContinuous2024, ongchocoDidThatJust2019}. 
Together, these findings suggest that memory formation is not constant across time but varies with the phase of state dynamics.
This pattern can be understood through the alternation between segregation and integration described in \ref{Neural State Transitions}.
Within a stable state, neural populations are relatively segregated, processing information within their specialized functions  \citep{shineDynamicsFunctionalBrain2016, tognoli_metastable_2014}.
At boundaries, integration is maximal, allowing the hippocampus to bind information across many cortical areas and providing a natural window for encoding, while simultaneously clearing the accumulated contents of working memory and compartmentalising the preceding event.
The temporal structure of our memories reflects the temporal structure of our neural state dynamics.
We remember experiences as discrete episodes with beginnings, middles and ends; and regularly over- or underestimate the actual time passed \citep{lositskyNeuralPatternChange2016a,ezzyatSimilarityBreedsProximity2014,bangertCrossingEventBoundaries2020,ongchocoEventSegmentationStructures2023,clewettPupillinkedArousalSignals2020}, something one might assume is because that is how our brains naturally segment the continuous flow of experience.

Similar to the nested nature of neural state during perception, memory may be organized in a nested structure. 
A short state transition in sensory cortex might trigger encoding of a perceptual detail, while a longer state transition in prefrontal cortex might trigger encoding of a higher-level goal or context.
The hippocampus can bind across these levels, creating memories that preserve the hierarchical structure of the original experience.
This nested organisation of memory mirrors the hierarchical process memory architecture described by \citet{hassonHierarchicalProcessMemory2015}.
Just as cortical areas accumulate context at different timescales during ongoing processing, the resulting memory traces retain this multi-scale temporal structure.

\section{Synthesis} 

To bring together the different elements we have discussed in the previous sections, we first present a number of core principles and then provide an example of how they could shape the functioning of our brain in an everyday situation. 

\subsection{Core Principles} 

\begin{enumerate}

\item \textbf{Temporal hierarchy mirrors spatial hierarchy.}
Just as perceptual representations become increasingly abstract along the cortical hierarchy, so too do their neural states become increasingly prolonged.
Primary sensory regions exhibit brief neural states reflecting fine-grained stimulus features, while association cortices maintain longer neural states that span the duration of meaningful events.
This is the case by and large, though exceptions do exists and should be acknowledged.

\item \textbf{Boundaries are moments of network reconfiguration across systems; stable periods are moments of segregation.}
Within a neural state, processing is relatively modular where specialized circuits operate semi-autonomously on their respective computations.
At state boundaries, functional connectivity reconfigures, enabling distant regions exert more influence on one another, temporarily increasing the capacity for information integration across systems.
This alternation between segregation and integration may represent the brain's solution to the competing demands of specialised processing and coherent behaviour.

\item \textbf{Neural states are the substrates of generative models.}
The content of a neural state is not simply a reflection of current input but an active model's encoding thorough which expected patterns of activity are generated across the processing hierarchy, built from accumulated past context.
State transitions are triggered when the model can no longer adequately account for incoming information, prompting a reconfiguration that requires heightened information integration and a new neural state.
This generative character explains why familiar events are segmented differently than novel ones, and why neural state boundaries shift earlier in time with increasing expertise.

\item \textbf{Memory is structured by state dynamics.}
The hippocampus exploits state boundaries as natural moments for encoding, capturing the distributed activity patterns that characterize an event just as it concludes.
The states themselves serve as the representational content of memory, replayed during consolidation and reinstated during retrieval.
The temporal structure of our memories reflects the temporal structure of our neural state dynamics.

\item \textbf{The nested hierarchy operates bidirectionally.}
Longer-duration states in higher-order regions constrain the space of possible states in faster-operating regions.
At the same time, states in faster operating regions constrain the states in slower operating regions.
For example, in vision, your goal state biases which object-level representations become active, but incoming object-level representations also bias what goals are possible.
These bidirectional constraints provide the mechanism through which context shapes perception and action and vice versa.
\end{enumerate}

\subsection{How neural states shape everyday information processing}

Having explored and contextualised these core principles, in this section we would like to illustrate the functioning of our brain in an everyday situation.
Consider the simple everyday scenario of trying to come up with a simple everyday scenario for a scientific paper.
This experience, lasting perhaps around an hour, unfolds through a cascade of nested neural states operating across multiple spatial and temporal scales.
As you scan the room for inspiration, your visual system initiates a rapid sequence of saccade-coupled states in primary visual cortex, each lasting roughly 200-300 milliseconds, sampling the layout of the desk, art on the walls, and books scattered around you.
These micro-states nest within longer object-level representations like the keyboard, a notebook and some pens, a specific painting, each maintaining stable patterns in ventral stream areas for a few seconds.
These in turn nest within a scene-level state in parahippocampal and retrosplenial cortices that persist for tens of seconds as you survey the room.
At the apex of this hierarchy, higher order areas like prefrontal and default mode network nodes maintain overarching situation model states, like 'write the first draft of this paper', which span minutes and longer and guide your behaviour.

When you receive an unexpected call from a friend, prediction error(s) cascade through the hierarchy.
Scene-level states transition as your spatial attention shifts.
You answering the call involves a sequence of motor action coupled states in motor, premotor, and sensory regions, again nested within a broader goal state such as `connect your headphone and sit down'.
Throughout, your hippocampus monitors these states and boundaries to bind the distributed patterns across the cortex into a coherent episodic trace:
`My friend called me while I was working'.

During your conversation the same principles apply.
Your friends speech unfolds through phoneme-level states (around tens of milliseconds), word-level states (around hundreds of milliseconds), sentence-level states (around seconds), and topic-level states (minutes and more).
At each of these levels, the accumulated process memory within the current state shapes the generative model, enabling you to anticipate upcoming words, follow the argument, and maintain the conversational context.
When your friend delivers surprising news, the resulting prediction error triggers state transitions across multiple scales.
This moment between not-knowing and knowing this news will likely be encoded with particular strength, as the hippocampus exploits the boundary to bind the distributed activity patterns into a lasting memory trace.

Meanwhile, slower whole-brain states modulate this entire process.
Your arousal levels, whether you are stressed or relaxed, influences the pool of potential states, as well as shapes the transition probabilities between states, affecting how quickly you respond, how deeply you encode the experience, or what comes to your mind during the conversation.

This example illustrates the core thesis of this review.
The cognitive framework of Event Segmentation and the mechanistic concept of Metastable Neural Activity describe the same underlying phenomenon from complementary perspectives.
ES provides the cognitive architecture, explaining how the brain segments experience (for comprehension, prediction, memory) and what these segments represent (hierarchically organized events from actions to episodes).
MNA provides the neural implementation models, explaining how segmentation emerges from the brain's intrinsic dynamics (through communication through coherence, attractor competition, functional coupling, and metastability) and what neural signatures accompany it (stable activity patterns punctuated by brief reconfiguration windows).

\section{Future Research}
The core principles we have distilled, generate specific, testable predictions that can guide future research.
Rather than treating event segmentation and metastable dynamics as separate phenomena to be studied in isolation, this synthesis encourages investigations that span cognitive and neural levels of analysis, and that examine how dynamics at different scales interact.
We outline several research directions that emerge naturally from these principles:

\paragraph{Cross-scale coupling mechanisms}
How exactly do neural states at different temporal scales interact?
While we have evidence that slower states constrain faster ones \citep{honey_slow_2012,hassonHierarchicalProcessMemory2015,geerligs_partially_2022,baldassano_discovering_2017,gozukaraMultiScaleAntiCorrelatedNeural2026}, the precise mechanisms remain unclear.
How do longer states modulate the attractor landscape in faster regions?
What is the role of subcortical circuitry in gating this information flow?
What cognitive consequences does this mechanism have?
Simultaneous recordings at multiple spatial scales, combined with computational modelling, could address these questions.

\paragraph{Slow oscillations}
This perspective could lead us to reconsider how we think about neural oscillations.
We typically investigate oscillations in the range of 0.5 to 80 Hz \citep{buzsakiRhythmsBrain2006,buzsakiNeuronalOscillationsCortical2004,cohenAnalyzingNeuralTime2014}.
However, most behaviourally relevant, and even conscious neural and mental phenomena occur at speeds slower than 0.5 Hz; and if there is good reason to suspect that we could apply our extensive knowledge at the faster scales to learn more about the slower ones. 
There is already evidence that these slower dynamics are functionally meaningful.
Using electrocorticographic (ECoG) recordings during movie viewing, \citet{honey_slow_2012} demonstrated that slow ($<$0.1 Hz) fluctuations of broadband high-frequency power are disproportionately expressed in higher-order cortical regions with long temporal receptive windows, are reliably time-locked to naturalistic stimuli, and become significantly more reliable when stimuli contain intact long-range temporal structure compared to scrambled controls.
Yet these slow dynamics remain largely disconnected from the oscillatory frameworks that have proven so productive at faster timescales; principles like communication through coherence and cross-frequency coupling offer mechanistic accounts of how neural populations coordinate and exchange information through phase relationships, but it remains unclear whether analogous coupling mechanisms operate between the slow fluctuations that govern neural state dynamics and the faster oscillations nested within them.
Investigating whether such principles extend across the brain's full range of temporal scales could yield a more unified mechanistic account of how metastable states emerge and interact across the cortical hierarchy.

\paragraph{The generative model at state boundaries}
If neural states reflect underlying generative models, several questions follow about what happens at their boundaries.
What determines the threshold at which a model is abandoned in favour of a new one?
Is this threshold adaptive, changing with task demands or uncertainty?
Can we observe signals of generative model failure that precede state transitions, and do these signals differ across levels of the hierarchy?
The distinction between surprisal and Bayesian surprise as drivers of state boundaries also warrants direct empirical comparison across different sensory modalities and task contexts.

\paragraph{Within-state accumulation dynamics}
The proposal that generative models are refined through the continuous accumulation of context within a state generates specific predictions.
Prediction accuracy should increase as a function of time spent within a stable state, and this improvement should be more pronounced at higher levels of the hierarchy where longer timescales allow for more extensive accumulation.
Combining neural state detection methods with time-resolved measures of prediction (e.g., language model surprisal, anticipatory neural signals) could test this directly.

\paragraph{Individual differences in segmentation dynamics}
People vary in their event segmentation abilities, with downstream consequences for memory and comprehension \citep{sargent_event_2013,kurby_segmentation_2008,sava-segal_individual_2023,baileyActionPerceptionPredicts2013}.
How do these individual differences reflect differences in metastable dynamics?
Clinical populations with known segmentation deficits (e.g., Parkinson's disease, ADHD) offer natural experiments for testing these relationships.

\paragraph{The role of hippocampal-cortical dialogue}
The hippocampus appears to both monitor and influence cortical state dynamics, encoding memories at neural state boundaries and potentially triggering state transitions during retrieval \citep{ben-yakov_hippocampal_2018,baldassano_discovering_2017,hahamy_human_2022,luNeuralNetworkModel2022a,pachecoestefanCoordinatedRepresentationalReinstatement2019,michelmannFasttimescaleHippocampalProcesses2025,dankerTrialbyTrialHippocampalEncoding2017,lee_anticipation_2021}.
What determines when the hippocampus `listens' versus when it `speaks'?
How does this dialogue unfold across the nested hierarchy?
The phenomenon of anticipatory reinstatement, where memory-driven predictions trigger state transitions before the expected stimulus, suggests that the hippocampus can actively shape cortical generative models, but the conditions and mechanisms of this influence remain to be characterised.

\paragraph{Training and expertise}
Expertise in a domain (music, sport, language) is associated with different event segmentation patterns \citep{blasingSegmentationDanceMovement2015,levineGoalBiasAction2017, nobleEventSegmentationBiological2014,newberryDoesSemanticKnowledge2019}.
Does this reflect changes in metastable dynamics, perhaps like more efficient state transitions, or better-tuned predictive models?
Can training interventions that target segmentation abilities improve broader cognitive outcomes?

\paragraph{State dynamics in action and social interaction}
Most research on neural states has used passive paradigms (e.g. movie-watching, or rest).
How do state dynamics operate when we are actively engaging with the environment, making decisions, and interacting with others?

\paragraph{Temporal brain reading/writing targets}
If neural state boundaries are windows of heightened integration, they may represent privileged moments for external intervention suggesting that the timing of brain stimulation relative to state dynamics is as important as its spatial targeting \citep{acero-pousaWhereWhenHow2026,bergmannBrainStateDependentBrain2018,zrennerClosedLoopNeuroscienceNonInvasive2016,decoAwakeningPredictingExternal2019}.
Can stimulation delivered at state boundaries more effectively modulate memory encoding, attentional reallocation, or predictive model updating than stimulation delivered mid-state?
Closed-loop systems that decode brain states in real time have already shown promise for rescuing memory \citep{ezzyatClosedloopStimulationTemporal2018} and enhancing hippocampal connectivity \citep{kragelClosedloopControlTheta2025}, but these approaches have not yet been informed by the nested hierarchy of metastable states described here.
More broadly, real-time decoding of state dynamics at multiple timescales could provide natural markers for brain-computer interfaces and neurofeedback, from sub-second motor states to slow whole-brain arousal fluctuations.

\paragraph{} By framing these questions we hope to encourage research that bridges traditionally separate literatures and levels of analysis.

\section{Conclusion}

In this review, we have argued that Event Segmentation and Metastable Neural Activity, two bodies of literature that have largely developed in isolation, describe the same underlying phenomenon: the organisation of neural activity into temporarily stable states that serve as the fundamental units of cognition.
ES provides the cognitive architecture, describing how experience is segmented into hierarchically organised events that support comprehension, prediction, and memory.
MNA provides the mechanistic account, describing how metastable dynamics, attractor competition, and oscillation-driven coupling give rise to these states across spatial and temporal scales.

Bringing these literatures together, we have proposed that metastable neural states reflect underlying generative models: active, context-accumulating internal models that generate expected patterns of activity across the processing hierarchy.
This framing provides a unifying thread across perception, decision making, action, prediction, and memory.
States persist while the generative model adequately accounts for incoming information; boundaries mark moments of model failure and reconfiguration, accompanied by heightened integration, and memory encoding.
The accumulated process memory within a state is simultaneously what enables its predictions and what constitutes its mnemonic trace, dissolving the traditional boundary between prediction and memory.

Understanding how neural states emerge, interact, and shape cognition brings us closer to understanding the brain in its natural mode of operation.

\section{References}
\begingroup
\sloppy
\printbibliography
\endgroup
\end{document}